\documentclass[11pt]{article}
\usepackage{float} 
\usepackage[symbol]{footmisc} 
\usepackage[superscript,sort]{cite}
\usepackage{array}
\usepackage{bm}

\usepackage[normalem]{ulem} 
\usepackage{amsmath,amssymb}
\usepackage{amsthm,amscd,amsxtra,amsfonts,amsmath,amssymb,multirow}
\usepackage{wrapfig}
\usepackage[footnotesize]{caption}
\usepackage{subcaption}
\usepackage{threeparttable} 
\usepackage{helvet}

\usepackage{booktabs}
\usepackage[table]{xcolor}
\usepackage{xcolor,colortbl}
\usepackage{placeins}
\usepackage{graphicx}
\usepackage[T1]{fontenc}
\usepackage[linesnumbered,ruled]{algorithm2e} 
\definecolor{MyPurple}{rgb}{1,0,1}

\newcommand{\beq}[1]{\begin{equation} \label{#1}}
\newcommand{\eeq}{\end{equation}}
\newcommand{\barray}{\begin{array}{ll}}
\newcommand{\earray}{\end{array}}

\usepackage{longtable} 
\usepackage[title]{appendix}

\DeclareMathOperator*{\argmin}{arg\,min}

\begin{document}
\pagenumbering{roman}

\clearpage \pagebreak \setcounter{page}{1}
\renewcommand{\thepage}{{\arabic{page}}}

\title{Persistent cohomology for data with multicomponent heterogeneous information}

\author{
Zixuan Cang$^1$, 
 and
Guo-Wei Wei$^{1,2,3}$ \footnote{
wei@math.msu.edu
}\\
$^1$ Department of Mathematics \\
$^2$  Department of Biochemistry and Molecular Biology\\
$^3$ Department of Electrical and Computer Engineering \\
Michigan State University, MI 48824, USA \\
}

\date{}
\maketitle
\maketitle
 \begin{abstract}  
Persistent homology is a powerful tool for characterizing the topology of a data set at various geometric scales. When applied to the description of molecular structures, persistent homology can capture the multiscale geometric features and   reveal certain interaction patterns in terms of topological invariants. However, in addition to the geometric information, there is a wide variety of non-geometric information of molecular structures, such as element types, atomic partial charges, atomic pairwise interactions, and electrostatic potential function, that is not described by persistent homology. Although element specific homology and electrostatic persistent homology can encode some non-geometric information  into geometry based topological invariants, it is desirable to have a mathematical framework to systematically embed both geometric and non-geometric information, i.e.,  multicomponent heterogeneous information, into unified topological descriptions. To this end, we propose a mathematical framework  based on persistent cohomology. In our framework, non-geometric information can be either distributed globally or resided locally on the datasets in the geometric sense and can be properly defined on  topological spaces, i.e., simplicial complexes.  Using the proposed persistent cohomology based framework, enriched barcodes are extracted from datasets to represent  heterogeneous information. We consider a variety of datasets to validate the present formulation and illustrate the usefulness of the proposed persistent cohomology. It is found that the proposed framework using cohomology boosts the performance of persistent homology based methods in the protein-ligand binding affinity prediction on massive biomolecular datasets.

 \end{abstract}

{\setcounter{tocdepth}{5} \tableofcontents}

\section{Introduction}
With the advancements in sensor hardware, data collection software,  data organization and storage frameworks, various datasets are expanding in an unprecedented speed where a large part of the newly accumulated data is high-dimensional,  highly complex, diverse, and often noisy. The rapid growth of datasets demands robust and  automatic data analysis tools. However, many widely used data analysis methods make assumptions of data complexity and/or the underlying dimensionality. Some other methods often require knowledge from domain experts. An emerging family of data analysis methods called topological data analysis  (TDA) answer these demands by combining ideas from algebraic topology with other mathematical tools \cite{Carlsson:2009}.  Making minimal assumptions of data, TDA characterizes the shapes of data in various dimensions, scans over a wide range of scales and is often robust against noise. 

Computational homology represents the topological structures of various dimensions by algebraic structures usually based on a fixed discrete topological space of interest. A topological space of discrete representation can be derived by building a simplicial complex upon a point cloud with a chosen scale parameter determining the topology or by building a cubicial complex upon volumetric data with a chosen isovalue. Continuous topological spaces can also be approximated by discrete representations, e.g., a tessellation of a manifold. Homology groups contain generators associated with the holes of certain dimensions in the topological space. While computational homology  captures the shape characteristics of a fixed structure, such a characteristics is insufficient in describing the structure in detail. A great variety of other structures might share the same characteristics. In mathematics, it is common to eliminate degeneracy
by introducing an extra dimension. Therefore, instead of examining the data at a fixed scale, persistent homology scans a sequence of topological spaces associated with a varying parameter that determines the topologies built upon the data.

Persistence describes the shape and scale of data by representing the data as a continuum of topological spaces which is called a filtration and tracking the homology features along this course of varying spaces. In practice, the filtration is often constructed by ordering and connecting. Via filtration, a collection of topological spaces is built on the data associated with different values of a scale parameter. Persistent homology tracks at what stage of the filtration do homology generators appear and how they persist along the subsequent course of the filtration. The persistent homology theory was formulated along with practical algorithms by Edelsbrunner \textit{et al} \cite{Edelsbrunner01topologicalpersistence}. Formal algebraic foundation was later established by Zomorodian and Carlsson \cite{Zomorodian:2005}. An earlier work named size function \cite{Frosini:1999} examines the connected components of topological spaces and is actually a version of $0$th dimensional persistent homology. Persistent homology has found its applications in many fields, for example, image processing \cite{Bendich:2010}, biology \cite{dabaghian2012topological,chan2013topology,KLXia:2014c,KLXia:2015b,KLXia:2015c, Perea:2015b,protein,ZXCang:2017c}, and fields in mathematics such as dynamical systems \cite{fluidP}.
Theoretical development has flourished since persistent homology was proposed such as Zigzag persistence \cite{zigzag} and multidimensional persistence \cite{CZ09} and there has been continuous advancement in algorithm development such as Perseus \cite{mischaikow:nanda}, PHAT \cite{bauer2017phat}, and Ripser \cite{bauer2017ripser }, paving the way for the analysis of complex and large datasets.

Given a point cloud dataset, it can be regarded as embedded in the Euclidean space which allows the usage of radius filtration associated with alpha complex \cite{Alpha}. A more general distance filtration associated with a Vietoris-Rips complex \cite{hausmann1995vietoris} or a \v{C}ech complex can be used to allow a predefined distance function suitable for specific applications \cite{KLXia:2014c,KLXia:2015b}. It is also possible to use a more flexible construction by directly assigning filtration values to  simplices in a complex which is considered as the final structure at the end of the filtration. In many applications, persistent homology is used to analyze the topological structures of datasets with generalized but homogenous information. For example, once the genetic distance between genes is defined by the number of mutations, persistent homology can be used to analyze the topological property of a gene evolution dataset.   
When the information of a dataset is heterogeneous, i.e., it involves multicomponent  information, for which special treatments are needed. For example, vineyards \cite{Cohen:2006,Munch:2013} is used to study spatiotemporal data. 

In dealing with chemical and biological datasets,  persistent homology was found to neglect crucial chemical and/or biological information during the topological simplification of the geometric complexity. 
Element specific persistent homology was introduced to embed  chemical, physical and biological information in  molecular datasets into topological invariants \cite{ZXCang:2017a,ZXCang:2017b, ZXCang:2017c, ZXCang:2018a}. 
When persistent homology is applied to complex molecular structures, in addition to the point cloud in the Euclidean space representing the coordinates of atoms, there are additional physical and chemical information such as element types, atomic partial charges, Coulomb and van der Waals interactions between atoms, and hydrophobic interactions among carbon atoms.  An interesting question is how to elegantly encode physical and chemical information into topological representations. 
A more general situation is that the data has multiple dimensions with heterogeneous meanings and persistent homology analysis is done only within  certain dimensions. This leads to a general question of how to systematically incorporate  the information of other dimensions into topological analysis. Therefore, there is a natural need to integrate multicomponent heterogeneous information into topological representations. Although one can resort to the tight representative cycles of  homology generators \cite{obayashi2017volume}, we prefer the cohomology framework because it is flexible and it is natural to view cochains as assigning weights on the simplicial complex which provides more quantitative representations. We consider cohomology theory with a graph Laplacian or a Laplacian defined on simplicial complexes to localize and smooth the representatives of (co)homology generators in the data and describe the additional information in the form of cochains (functions on chains) by computing inner product of these cochains and the smoothed (persistent) cohomology representatives.

Cohomology provides a richer algebraic structure  for a topological space. The cohomology construction used in this work dualizes homology by building cochains as functions on chains in homology theory. Cohomology theory has been applied in both mathematics and the field of data analysis. One well known cohomology theory is the de Rham cohomology which studies the topological features of smooth manifolds using differential forms. The de Rham cohomology has led to further theoretical developments such as the Hodge theory. Recently, a discrete exterior calculus framework has been established \cite{hirani2003discrete} where  manifolds are approximated by mesh-like simplicial complexes and the discrete counterparts of the continuous concepts such as differential forms are defined thereafter. This framework has many applications, for example, the harmonic component of the discrete Hodge decomposition has been used in sensor network coverage problem to localize holes in a sensor network \cite{bell2012pydec}. Cohomology theory has also been applied to persistent homology analysis. A $1$-dimensional cohomology was used to assign circular values to the input data associated to a homology generator \cite{de2011persistent} which further led to applications in several fields including the analysis of neural data \cite{rybakken2017decoding} and the study of periodic motion \cite{vejdemo2015cohomological}. Persistent cohomology in higher dimensions has been used to produce coordinate representations which reduces dimensionality while retaining the topological property of data \cite{perea2018multiscale}.  Weighted (co)homology and weighted Laplacian were introduced with biological applications \cite{wu2018weighted}.  Computationally, the duality between homology and cohomology \cite{de2011dualities} has set the basis for constructing more efficient algorithms that utilize  cohomology to  compute persistent homology. Several code implementations, such as Dionysus \cite{Dionysus} and Ripser \cite{bauer2017ripser }, speed up the persistent homology computation by taking the advantage of this property.

In this work, we seek a formulation that  organizes geometric information into  simplicial complex while encoding chemical, physical and biological properties into functions fully or partially defined on simplicial complexes  locally associated with homology generators. To this end, we need a representation that can locate  homology generators. When manifold-like simplicial complexes are available, we can look for harmonic (in the sense of Laplace-de Rham operator) cohomologous cocycles under the framework of discrete exterior calculus \cite{hirani2010cohomologous}. A discrete version of the Hodge-de Rham theorem guarantees the uniqueness of the harmonic cocycle if certain conditions are satisfied \cite{hirani2010cohomologous}. However, this method requires the proper construction of the Hodge star operator which usually relies on a well-defined dual complex while in general applications, this is not always feasible. For example, when a user-defined distance matrix is used with the Rips complex, the distance may even not satisfy triangle inequality. Therefore, we relax our requirement on the accuracy of geometric localization  and represent the set of simplices of a certain dimension in a simplicial complex as a graph where the simplices are represented by graph nodes and their adjacency is treated as graph edges.  We can also define a Laplacian on simplicial complexes by first introducing an inner product of cochains and then constructing an adjoint of the coboundary operator. Then, the smoothness of a cocycle can be measured by a Laplacian. Specifically, given a representative cocycle of a homology generator, we look for a cohomologous cocycle that minimizes the norm of the output under the Laplacian. We can then consider such smoothed cocycles which distribute smoothly around the holes of certain dimensions as measures on  simplicial complexes and describe the input functions defined on the simplicial complexes by integrating with respect to these measures. The present formulation also utilizes a filtration process to assign a function over the filtration interval associated to each bar in the barcode representation to result in an enriched barcode representation of persistent cohomology. A modified Wasserstein distance is defined and implemented subsequently to facilitate the comparison of these enriched barcodes generated from datasets.

In the rest of this paper, basic background of persistent homology and cohomology is given in Section~\ref{sec:theory} and the persistent cohomology enriched barcodes with accompanying analysis tools are developed in Section~\ref{sec:method}. In Section~\ref{sec:results}, we illustrate the proposed method by  simple examples, example datasets, and the characterization of molecules. Finally,  we also demonstrate the utility of the proposed persistent cohomology by the prediction of  protein-ligand binding affinities from large datasets.  

\section{Motivation}
The development of the method in this work is for a scenario in topological data analysis, specifically persistent homology where the data has multiple heterogeneous dimensions while it is not appropriate to compute geometry/topology with all the dimensions. However, the dimensions that are not considered for topological characterization may carry useful information. This work re-acquire this additional information by building maps upon the persistent homology results using cohomology. 

More precisely, consider a dataset $X$ containing $N$ members each having an vector of size $\ell$ as description which can be stored as an $N\times \ell$ matrix. An example can be a set of $N$ atoms from a molecule with $\ell=3$ recording the Cartesian coordinates of the atoms where persistent homology computation can be directly done by considering the dataset as a point cloud in $\mathbb{R}^3$. However, there can be more information in the dataset. For example, in addition to the coordinates, there can be partial charges on the atoms. In this case, it is not appropriate to directly compute persistent homology by considering the dataset as a point cloud in $\mathbb{R}^4$. A more general situation is that the $\ell$ elements in the description vector contains both geometric information and non-geometric information. For simplicity, we assume the elements are already sorted such that the first $m$ are for geometry and the following $n$ elements are for non-geometric features. We can compute a certain dimensional persistent homology for the submatrix $X(1,\cdots,N;1,\cdots,m)$ and obtain a barcode $PH(X)=\{[b_i,d_i)\}_{i\in I}$ which is basically a collection of half-open intervals. Then, the method in this work derives a function $f^*:\mathbb{R}^2\times\mathbb{R}\rightarrow \mathbb{R}^n$ on the intervals reflecting the information given in $X(1,\cdots,N;m+1,\cdots,\ell)$. For a bar $[b_i,d_i)$ in the barcode and a filtration value (which is a parameter in persistent homology related to scale) $\epsilon\in [b_i,d_i)$, $f^*(b_i,d_i;\epsilon)$ describes the information carried by  $X(1,\cdots,N;m+1,\cdots,\ell)$ associating to this particular bar at the specific filtration value. In the earlier example with the $1$-dimensional persistent homology, this reflects the average charge of atoms contributing to the loop or tunnel associated to the bar.

\section{Theoretical background}\label{sec:theory}

\subsection{Simplex and simplicial complex}
For point clouds or atomic coordinates, their topological analysis and characterization can be carried out via simplices and simplicial complexes. A set of $k+1$ affinely independent points in $\mathbb{R}^n$ is a $k$-simplex denoted $\sigma$ which can be represented by $[v_0,\cdots,v_k]$ and each $v_i$ is called a vertex of the simplex. A simplex $\tau$ is a face of $\sigma$ if the vertices of $\tau$ form a subset of the vertices of $\sigma$ and is denoted $\tau\leq\sigma$.
A simplicial complex is a set of simplices which are convex hulls of affinely independent points. More specifically, a simplicial complex is a finite collection of simplices $X=\{\sigma_i\}_i$ satisfying that the intersection of any two simplices in $X$ is either empty set or a common face of the two and all the faces of a simplex in $X$ is also in $X$. The collection of all $k$-simplices in $X$ is denoted $X^k$. The dimension of a simplicial complex is the highest dimension of its simplices.


\subsection{Homology and cohomology}

Given a simplicial complex $X$, a $k$-chain on it is a finite formal sum of all simplices in $X^k$, $c=\sum\limits_i a_i\sigma_i$ with coefficients $a_i\in\mathbb{Z}_p$, the field of integers of modular $p$ where $p$  is a chosen prime number. The set of all $k$-chains in $X$ with the addition given by the addition of coefficients forms a group called the $k$th chain group denoted $C_k(X)$. The orientation of a simplex is given by the ordering of its vertices and two orderings give the same orientation if and only if they differ by an even number of permutations. For example, $[v_0,v_1]=-[v_1,v_0]$ and $[v_0,v_1,v_2]=[v_1,v_2,v_0]$. The boundary operator $\partial_k: C_k(X)\rightarrow C_{k-1}(X)$ is a linear mapping that maps a $k$-simplex to its codimensional $1$ faces,
\begin{equation*}
\partial_k([v_0,\cdots,v_k]) = \sum\limits_{i=0}^k(-1)^i[v_0,\cdots ,\widehat{v}_i,\cdots ,v_k],
\end{equation*}
where $\widehat{v}_i$ denotes the absence of $v_i$. When there is no ambiguity, we simply denote $\partial_k$ by $\partial$. We say that $k$-chain is a boundary if it is in the image of $\partial_{k+1}$. A $k$-chain is a $k$-cycle if its image under $\partial_k$ is $0$, the null set. The $k$th homology group is the quotient group $H_k(X)=\mathrm{Ker}(\partial_{k})/\mathrm{Im}(\partial_{k+1})$ containing equivalence classes of $k$-cycles. $\mathrm{Im}(\partial_{k+1})$ is a subgroup of $\mathrm{Ker}(\partial_k)$ following that $\partial_k\circ\partial_{k+1}=0$. Two $k$-cycles are in the same equivalence class in $H_k(X)$ if they differ by the boundary of a $(k+1)$-chain and they are called homologous. When there is no ambiguity, we simply use $H_k(X)$ for $H_k(X, \mathbb{Z}_p)$.

Cohomology is also a sequence of abelian groups associated to topological space $X$ and is  defined from a cochain complex, which  is a function on the chain group of the homology theory. Specifically,  a $k$-cochain is a function $\alpha: X^k\rightarrow R$ where $R$ is a commutative ring. The set of all $k$-cochains following the addition in $R$ is called the $k$th cochain group denoted $C^k(X,R)$. The coboundary operator $d_k: C^k(X,R)\rightarrow C^{k+1}(X,R)$ maps a cochain to a cochain one dimension higher and is the counterpart of boundary operators for chains, namely
\begin{equation*}
d_k(\alpha)([v_0,\cdots,v_k]) = \sum\limits_{i=0}^k(-1)^i\alpha([v_0,\cdots,\widehat{v}_i,\cdots,v_k]),
\end{equation*}
for a $k$-cochain $\alpha$. It should be noted that in the matrix representation of the two operators, $d_k$ and $\partial_{k+1}$ are transpose to each other. When there is no ambiguity, we simply refer to $d_k$ using $d$. A $k$-cochain is called a coboundary if it is in the image of $d_{k-1}$. A $k$-cochain is called a cocycle if its image under $d_k$ is $0$. The coboundary operators have the property that $d_k\circ d_{k-1}=0$ following that $d_k\circ d_{k-1}=\partial_{k+1}^T\circ\partial_k^T=(\partial_k\circ\partial_{k+1})^T$. The $k$th cohomology group is defined to be the quotient group $H^k(X,R)=\mathrm{Ker}(d_k)/\mathrm{Im}(d_{k-1})$. Two cocycles are called cohomologous if they defer by a coboundary.

In practice, some field is used instead of a ring due to the computation of persistent (co)homology. From now on, we consider finite fields $\mathbb{Z}_p$ with some prime $p$.

\subsection{Persistence}

We are interested in the evolution of a simplicial complex and hope to track how topological changes as the simplicial complex changes. Given a simplicial complex $X$ and a function $g: X\rightarrow \mathbb{R}$. For any $x\in\mathbb{R}$, a sublevelset of $X$ is defined as 
\begin{equation*}
X(x)=\{\sigma\in X| g(\sigma)\leq x\}.
\end{equation*}
The function $g$ is required to satisfy that $g(\tau)\leq g(\sigma)$ for any $\sigma$ and any $\tau\leq\sigma$.
Since $X$ is a finite collection of simplices, we can have a finite sorted range of $g$ as $\{x_i\}_{i=0}^l$ where $x_i<x_j$ if $i<j$. The filtration of $X$ associated with $g$ is an order sequence of subcomplexes of $X$,
\begin{equation}\label{eq:filtration}
\emptyset\subset X(x_0)\subset X(x_1)\subset\cdots\subset X(x_l)=X.
\end{equation}

We choose $\mathbb{Z}_p$ to be the coefficient field for chain groups. Persistent homology keeps tracks of the appearance and disappearance of homology classes along the filtration which also includes the information of homology of each fixed simplicial complex in filtration $\{X(x_i)\}_i$. Since we are working on a field, the homology groups $H_k(X(x_i))$ can be represented as vector spaces. The inclusion map connecting the groups induces a sequence of linear transformations on the vector spaces as
\begin{equation}
H_k(X(x_0))\rightarrow H_k(X(x_1))\rightarrow \cdots \rightarrow H_k(X(x_l)).
\end{equation}
A persistence module $\{V_i, \phi_i\}$ is a collection of a sequence of vector spaces $V_i$ and linear transformations connecting them $\phi_i: V_i\rightarrow V_{i+1}$. An interval module $\mathbb{I}_{[b,d)}$ is a special case of a persistence module with identity maps 
\begin{equation}
I_0\rightarrow I_2 \rightarrow \cdots \rightarrow I_l,
\end{equation}
where $I_i=\mathbb{Z}_p$ for $b\leq i<d$, and $I_i=0$ otherwise. 
A special case of theorem of Gabriel implies that a nice enough persistence module can be decomposed uniquely as a direct sum of interval modules, $\bigoplus_{[b,d)\in B}\mathbb{I}_{[b,d)}$. The collection of half open intervals $B$ can be visualized as barcodes, which represent topological invariants as horizontal bars,  or persistence diagrams, which  use dots in a 2D picture to describe topological events.  

Similarly, the persistent cohomology can be derived with the following relationship
\begin{equation*}
H^k(X(x_0), \mathbb{Z}_p)\leftarrow H^k(X(x_1), \mathbb{Z}_p)\leftarrow \cdots \leftarrow H^k(X(x_l), \mathbb{Z}_p).
\end{equation*}
The universal coefficient theorem for cohomology (Theorem 3.2 \cite{Hatcher:2001}) implies that there is a natural isomorphism $H^k(X,\mathbb{Z}_p)\equiv \mathrm{Hom}_{\mathbb{Z}_p}(H_k(X,\mathbb{Z}_p), \mathbb{Z}_p)$ so that the cohomology group can be considered as the dual space of the homology group. This property further implies that $\mathrm{rank}(H^k(X,\mathbb{Z}_p))=\mathrm{rank}(H_k(X,\mathbb{Z}_p))$ and thus persistent homology and persistent cohomology have identical barcodes \cite{de2011dualities}. 

\section{Method}\label{sec:method}
\subsection{Smoothed cocycle}

Some representative cocycles in persistent cohomology may not reflect the overall location and structure associated with their cohomology generators. To better embed the additional information in the data into cohomology generators, we look for a smoothed representative cocycle in each cohomology class. The smoothness of functions can usually be measured by a Laplacian. We construct smoothed representative cocycles with a Laplacian in this section.

\subsubsection*{Laplacian on simplicial complex}
A Laplacian for cochains can be defined by first defining an inner product and using the induced adjoint operator. Assuming the case of real number,  for $\alpha_1,\alpha_2\in C^k(X,\mathbb{R})$, the inner product can be defined as
\begin{equation}
<\alpha_1, \alpha_2>_k = \sum\limits_{\sigma\in X^k}\alpha_1(\sigma)\alpha_2(\sigma).
\end{equation}
Then, the adjoint $d_k^*: C^{k+1}(X,\mathbb{R})\rightarrow C^k(X,\mathbb{R})$ of the operator $d_k$ with respect to this inner product can be defined by
\begin{equation}
<d_k\alpha,\beta>_{k+1} = <\alpha,d_k^*\beta>_k,\,\mathrm{for}\,\alpha\in C^{k}(X,\mathbb{R}),\beta\in C^{k+1}(X,\mathbb{R}).
\end{equation}
Weights reflecting size of simplices can be used to reflect the geometry by defining a weighted inner product,
\begin{equation}
<\alpha_1, \alpha_2>_k^w = \sum\limits_{\sigma\in X^k}s_\sigma\alpha_1(\sigma)\alpha_2(\sigma),
\end{equation}
where $s_\sigma$ is the size of $\sigma$ such as area or volume if $\sigma$ is a $2$- or $3$-simplex.
Then, a Laplacian on $X^k$ can be defined by
\begin{equation}\label{eq:sclaplacian}
\mathcal{L}_{sc} = d_{k}^*d_k + d_{k-1}d_{k-1}^*.
\end{equation}
An inner product based on Wedge product can also be constructed if a manifold like simplicial complex is given.

\subsubsection*{Weighted graph Laplacian}
We can also represent $X^k$ as a graph where the nodes are simplices and the edges describe adjacency. Consider a graph associated to $X^k$ where each $k$-simplex is represented by a node and there is an edge if two $k$-simplices have nonempty intersection. Note that this is a simple graph and we define a weight matrix $W=(w_{ij})$ to be 
\begin{equation}\label{eq:weightmatrix}
w_{ij} = 
\begin{cases}
v(\sigma_i) v(\sigma_j),\, \sigma_i\cap\sigma_j\neq\emptyset,\\
0,\, \mathrm{otherwise},
\end{cases}
\end{equation}
where $v(\sigma)$ is the size of $\sigma$, for example, the area of $2$-simplices and the volume of $3$-simplices. The size of a $0$-simplex is defined to be $1$.
Then, the $W$-weighted graph Laplacian \cite{chung1997spectral} $\mathcal{L}_W$ is defined as
\begin{equation}\label{eq:wlaplacian}
{\mathcal{L}_W}_{i,j} =
\begin{cases}
1 - w_{ij}/w_i, \, \mathrm{if}\, i=j,\, \mathrm{and}\, w_i\neq 0, \\
-w_{ij}/\sqrt{w_i w_j}, \, \mathrm{if}\, \sigma_i\cap\sigma_j\neq\emptyset, \,\mathrm{and}\, i\neq j, \\
0,\, \mathrm{otherwise},
\end{cases}
\end{equation}
where $w_i=\sum\limits_{j}w_{ji}$.
A $k$-cochain $\alpha$ can be represented by a column vector given the basis in Eq.~(\ref{eq:cochainbasis}).
The matrix $\mathcal{L}_W$ measures the difference between the value of the cochain on a simplex and the values on the neighbors of this simplex.  A large penalty is given to prevent rapid changes through smaller simplices.

\subsection{Persistent cohomology enriched  barcode }\label{sec:epb}

We describe the work-flow in this section. Given a simplicial complex $X$ of dimension $n$, and a function $f:X^k\rightarrow \mathbb{R}$ with $0\leq k \leq n$, we seek a method to embed the information of $f$ on the persistence barcodes obtained with a chosen filtration of $X$. In other words,  we seek a representation of $f$ on cohomology generators. To this end, smoothed representations are first computed for cohomology generators. One of such smoothed representations induces a measure on the simplicial complex which allows us to integrate $f$ on $X$. We describe the protocol of our approach  below.

\subsubsection*{Dimension greater than 0}
Consider a filtration of $X$, $\emptyset=X(x_0)\subseteq X(x_1) \subseteq\cdots\subseteq X(x_n)=X$ and an associated persistent cohomology  with a prime $p$ other than $2$.  Let $\omega$ be a representative cocycle for a persistence interval $[x_i,x_j)$ of dimension $k>0$. The cocycle $\omega$ is first lifted to a cocycle $\hat{\omega}$ with integer coefficients satisfying that $\omega(\sigma)\equiv \hat{\omega}(\sigma)\, (\mathrm{mod}\,p)$ and $\hat{\omega}(\sigma)\in \{i\in\mathbb{Z}: -(p-1)/2\leq i \leq (p-1)/2\}$ for all $\sigma\in X^k$. This is almost always possible \cite{de2011persistent}. Now that $\hat{\omega}$ is an interger cocycle and thus also a real cocycle. With a basis for $k$-cochains $\{\alpha_{\sigma_i}\}_i$ where 
\begin{equation}\label{eq:cochainbasis}
\alpha_{\sigma_i}(\sigma)=
\begin{cases}
1,\, \sigma=\sigma_i \\
0,\, \mathrm{otherwise}
\end{cases}
\end{equation}

Given a Laplacian on cochains $\mathcal{L}$ to measure smoothness, a smooth cocycle $\bar{\omega}$ can be obtained by solving a least square problem,
\begin{equation}\label{eq:optimization}
\bar{\alpha}=\argmin\limits_{\alpha\in C^{k-1}(X,\mathbb{R})}\|\mathcal{L}(\hat{\omega}+d\alpha)\|_2^2,
\end{equation}
and letting $\bar{\omega}=\hat{\omega}+d\bar{\alpha}$. This smoothed cocycle $\bar{\omega}$ induces a measure $\mu$ on $X^k$ by setting 

\begin{equation}\label{eq:measure}
\mu(\sigma)=|\bar{\omega}(\sigma)|.
\end{equation}

To obtain a sequence of such smoothed real $k$-cocycles for the cohomology generator along a persistence interval, we restrict the representative integer cocycle $\hat{\omega}$ to subcomplexes of $X$ and repeat the smoothing computation. Consider the integer $k$-cocycle $\hat{\omega}|_{X(x)}$ at filtration value $x$. The corresponding smoothed real $k$-cocycle $\bar{\omega}_x$ can be obtained by running the optimization problem for $\hat{\omega}|_{X(x)}$ as Eq.~(\ref{eq:optimization}) on $C^{k-1}(X(x),\mathbb{R})$ and it induces a measure $\mu_x$ on $X^k(x)$ as described in Eq.~(\ref{eq:measure}). It suffices to compute for all different filtration values in $[x_i,x_j)$ because we have a finite filtration which gives $\{\mu_{x_\ell}\}_{\ell=i}^{j-1}$.

A function of filtration values $f^*$ can be defined for each persistent interval $[x_i,x_j)$ as 
\begin{equation}\label{eq:enrichedfunc}
f^*(x) = \int_{X^k(x)}f d\mu_x\bigg/\int_{X^k(x)}d\mu_x
\end{equation}
for $x\in[x_i,x_j)$. We call each of the collection of persistent intervals being associated with one such function $f^*$ an enriched persistent barcode.

\subsubsection*{Dimension 0}
In the case of dimension $0$, persistent homology tracks the appearance and merging of connected components. It is convenient to assign a smooth $0$-cocycle to a persistent interval by assigning $1$ to the nodes in the connected component associated with the interval right before the generator is killed due to merging with another connected component. This is implemented with a union-find algorithm.

\subsection{Preprocessing of the input function}\label{sec:inputfunction}
When given the original input function associated with the input data, we first need to generate a cochain of the dimension of interest out of this input function. The procedures in several situations are discussed in the rest of this section.

\subsubsection*{Case 1}
When given a function $f_0:X^{k_0}\rightarrow \mathbb{R}$, and we are interested in its behavior associated with a $k$-dimensional homology where $k_0\neq k$. We need to interpolate or extrapolate $f_0$ to a function $f:X^k\rightarrow \mathbb{R}$. We assume that $f_0$ is unoriented, i.e. $f_0(\sigma)=f_0(-\sigma)$. A simple way is to take unweighted averages,
\begin{equation}\label{eq:Fa}
f_a(\sigma)=\frac{1}{n_\sigma}\sum\limits_{i=1}^{n_\sigma}f_0(\sigma'_i),
\end{equation}
where each $\sigma'_i$ is a $k_0$-simplex satisfying that $\sigma'_i<\sigma$ if $k> k_0$ and $\sigma'_i > \sigma$ if $k < k_0$ and $n_\sigma$ is the total number of such $k_0$-simplices.
A weighted version based on geometry can be defined as
\begin{equation}\label{eq:Fw}
f_w(\sigma)=\sum\limits_{i=1}^{n_\sigma}w_if_0(\sigma'_i)\bigg/\sum\limits_{i=1}^{n_\sigma}w_i,
\end{equation}
where $w_i$ is the reciprocal of the distance between the barycenters of $\sigma$ and $\sigma'_i$.

An example of this situation is the pairwise interaction strengths between atoms of a molecule which are naturally defined on edges connecting the vertices representing the atoms. Another example is the atomic partial charges defined on the vertices representing the atoms in a molecule or a molecular complex.

\subsubsection*{Case 2}
When given a function $f_0:\mathbb{R}^n\rightarrow\mathbb{R}$ with $n>= k$ and a geometric simplicial complex, we can integrate it on every $k$-simplex in $X$ to obtain a function $f_i:X^k\rightarrow\mathbb{R}$. For simplicity, we require $f_0$ to be bounded. Then, $f_i$ is defined as
\begin{equation}
f_i(\sigma) = \int_\sigma f_0d\sigma\bigg/\int_\sigma d\sigma,
\end{equation} 
for a $k$-simplex $\sigma$ and $\int_\sigma d\sigma$ computes the $k$-dimensional volume of $\sigma$. In many cases, $f_0$ is given as results of numerical simulations which is often defined on grid points. Then, the integrals can be computed by some chosen quadrature formula and interpolating $f_0$ to the collocation points.

\subsection{Modified Wasserstein distance for persistent cohomology enriched  barcodes}

An enriched bar can be represented by three elements, birth value  $b$,  death value $d$, and  function $f^*$ constructed by Eq.~(\ref{eq:enrichedfunc}). Given two enriched barcodes of the same dimension represented by $B=\{ \{ b_i,d_i,f^*_i\} \}_{i\in I}$ and $B'=\{ \{b'_j,d'_j,f'^*_j\} \}_{j\in J}$, we would like to compute their distance analogous to Wasserstein distance. We first define two pairwise distances, i.e.,  $\Delta_b$ that measures the distance between two persistence bars
\begin{equation}
\Delta_b\left([b,d),[b',d')\right) = \mathrm{max}\{|b-b'|,|d-d'|\}
\end{equation}
and   $\Delta_f$ that measures the distance between $f^*$ and $f'^*$
\begin{equation}\label{eq:fint}
\Delta_f(f^*, f'^*) = \left|\frac{1}{d-b}\int_{b}^{d}f^*(x)dx-\frac{1}{d'-b'}\int_{b'}^{d'}f'^*(x)dx\right|.
\end{equation}
In practice, it sometimes is too costly to compute the output values of $f^*$ for all possible filtration values, and only a subset of possible filtration values is selected, such as only the middle value of a bar.
In this case, we use the middle Riemann sum to approximate the integration in Eq.~(\ref{eq:fint}).
For a bijection $\theta:\bar{I}\rightarrow\bar{J}$ where $\bar{I}$ and $\bar{J}$ are subsets of $I$ and $J$, the associated penalties are defined as
\begin{equation}
\begin{aligned}
P_b(\theta;q,B,B') = &\sum\limits_{i\in \bar{I}}\left(\Delta_b\left([b_i,d_i),[b'_{\theta(i)},d'_{\theta(i)})\right)\right)^q\\
 &+ \sum\limits_{i\in I\setminus\bar{I}}\left(\Delta_b\left([b_i,d_i),[(b_i+d_i)/2,(b_i+d_i)/2)\right)\right)^q \\
&+ \sum\limits_{i\in J\setminus\bar{J}}\left(\Delta_b\left([b'_i,d'_i),[(b'_i+d'_i)/2,(b'_i+d'_i)/2)\right)\right)^q
\end{aligned}
\end{equation}
and 
\begin{equation}
\begin{aligned}
P_f(\theta;q,B,B') = &\sum\limits_{i\in \bar{I}}\left(\Delta'_f\left(f^*_i,f'^*_{\theta(i)}\right)\right)^q \\
&+\sum\limits_{i\in I\setminus\bar{I}}\left(\Delta'_f(f^*_i,0)\right)^q \\
&+\sum\limits_{i\in J\setminus\bar{J}}\left(\Delta'_f(f'^*_i,0)\right)^q.
\end{aligned}
\end{equation}
 The $q$th modified Wasserstein distance is defined as
\begin{equation}
W^{q,\gamma}(B,B') = \inf\limits_{\theta\in\Theta}\left(\gamma P_b(\theta;q,B,B')+(1-\gamma)P_f(\theta;q,B,B')\right)^{\frac{1}{q}},
\end{equation}
where $\gamma$ is a weight parameter and we denote the minimizer by $\theta^{q,\gamma}$. Similar to receiver operating characteristic curve, instead of fixing $\gamma$ we let it change from $0$ to $1$ which results in a function $\mathcal{W}^q:[0,1]\rightarrow\mathbb{R}^2$ defined as
\begin{equation}\label{eq:wassersteincurve}
\mathcal{W}^q(\gamma) = [P_b(\theta^{q,\gamma};q,B,B')^{\frac{1}{q}}, P_f(\theta^{q,\gamma};q,B,B')^{\frac{1}{q}}],
\end{equation}
and we call it a Wasserstein characteristic curve.

The optimization problem can be considered as an assignment problem and solved by Hungarian algorithm.
Given two enriched barcodes $B=\{\{b_i,d_i,f^*_i\}\}_{i=1}^m$ and $B'=\{\{b'_j,d'_j,f'^*_j\}\}_{j=1}^n$, we first construct pseudo barcode for each of them to account for the situation where a bar is not paired with another. The pseudo barcodes are $B_{B'}=\{\{(b'_j+d'_j)/2,(b'_j+d'_j)/2,0\}\}_{j=1}^n$ and $B'_{B}=\{\{(b_i+d_i)/2,(b_i+d_i)/2,0\}\}_{i=1}^m$. Then the assignment problem between $B\cup B_{B'}$ and $B'\cup B'_B$ is solved with the cost $(\gamma P_b+(1-\gamma)P_f)^{\frac{1}{q}}$. The linear\_sum\_assignment tool under optimize module of SciPy package \cite{scipy} is used.

\subsection{Implementation}
The adjoint of the coboundary operator required to construct the Laplacian on simplicial complex defined in Eq.~\ref{eq:sclaplacian} can be obtained by taking the transpose.
In this section, we focus on the implementation of enriched barcodes using weighted graph Laplacian.
The most costing steps of constructing persistent cohomology enriched barcodes are the construction of the weight matrix, and the subsequent least square optimization problem. To construct the weight matrix defined in Eq.~\ref{eq:weightmatrix}, we only need to go over all the simplices once as described in Algorithm~\ref{alg:weightmatrix}. Then, the weight matrix with respect to a certain filtration value is simply a submatrix and the weighted graph Laplacian can be formed following Eq.~\ref{eq:wlaplacian}.
\begin{algorithm}[ht]
    \SetKwInOut{Input}{input}\SetKwInOut{Output}{output}\SetKwInOut{Parameter}{parameter}
    
    \Input{The boundary matrix for the filtration, $B$ of size $n\times n$ with associated set of simplices $X=\{\sigma_i\}_{i=1}^n$.}
    \Output{A weight matrix for $k$-simplices, $W$ of size $m\times m$.}
	\Parameter{A vector $I$ of size $n$ mapping the index of simplices.\\ $NzId_i$, a vector of the column numbers of nonzero elements of row $i$ of $B$ with an ascending order.}   
    
    $j \leftarrow 1$ \;
    \For{$i \leftarrow n$ \KwTo $1$}{
        \uIf{$\vert\sigma_i\vert == k+1$}{
            $I[i]\leftarrow j$ \;
            $j++$ \;
        }
        \uElseIf{$\vert\sigma_i\vert \leq k$}{
        	\For{$\hat{i}\leftarrow 1$ \KwTo $length(NzId_i)-1$}{
        	    \For{$\hat{j}\leftarrow \hat{i}+1$ \KwTo $length(NzId_i)$}{
        	        $\tilde{i}\leftarrow I[NzId_i[\hat{i}]$, $\tilde{j}\leftarrow I[NzId_i[\hat{j}]$\;
        	        \If{$W[\tilde{i}, \tilde{j}] == 0$}{
        	        	$W[\tilde{i}, \tilde{j}],W[\tilde{j}, \tilde{i}] \leftarrow size(\sigma_{\tilde{i}})*size(\sigma_{\tilde{j}})$\;
        	        }
        	    }
        	}
        }
    }
\caption{An algorithm for constructing the weight matrix for the graph Laplacian}\label{alg:weightmatrix}
\end{algorithm}

Note that the weight matrix and the weighted graph Laplacian are usually sparse matrices. The system are implemented as sparse matrices and the least square optimization problem in Eq.~\ref{eq:optimization} is solved by using the sparse.linalg.lsqr module in SciPy package\cite{scipy}.

\FloatBarrier

\section{Numerical results}\label{sec:results}

\subsection{A minimalist example of persistent cohomology}
\begin{figure}[h]
\begin{center}
\includegraphics[width=0.25\textwidth]{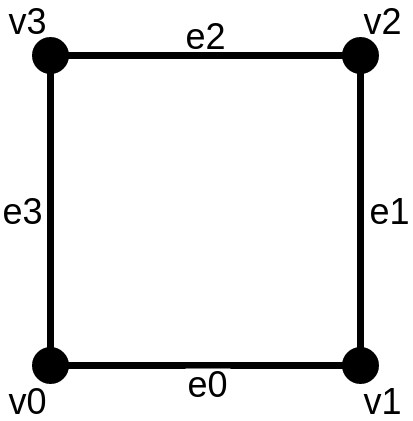}
\end{center}
\caption{A simple example}\label{fig:minimalistexample}
\end{figure}

Consider a simplicial complex $X$ with four vertices and four edges with unit length that forms a square as shown in Figure~\ref{fig:minimalistexample}. The $1$-cochain $\hat{\omega}=[1,0,0,0]^T$ is a real cocycle. The notation means that $\hat{\omega}(e0)=1$ and $\hat{\omega}(e1)=\hat{\omega}(e2)=\hat{\omega}(e3)=0$. The weighted Laplacian matrix $\mathcal{L}_W$ defined in Eq.~(\ref{eq:wlaplacian}) for $X^1$ is
\begin{equation*}
\frac{1}{3}
\begin{bmatrix}
2 & -1 & 0 & -1 \\
-1 & 2 & -1 & 0 \\
0 & -1 & 2 & -1 \\
-1 & 0 & -1 & 2
\end{bmatrix}
\end{equation*}
when a uniform weight of $1$ is assigned to all edges. Then, we obtain a smoothed cocycle $\bar{\omega}=\omega+d\bar{\alpha}=[0.5, 0.5, 0.5, 0.5]^T$ with a $0$-cochain $\bar{\alpha}=[1, 0.5 ,1, 1.5]^T$ which minimizes $\|\mathcal{L}_W(\hat{\omega}+d\bar{\alpha})\|_2^2$ to $0$.

\FloatBarrier


\subsection{Persistent cohomology analysis of two simple datasets}
In this section, we show the smoothed representative $1$- and $2$-cocycles and the enriched barcodes using artificial datasets. We create some example input functions defined on the nodes and aim to reflect the information about these functions on the enriched barcodes.

\subsubsection*{Two adjacent annuluses}
\begin{figure}[h]
\begin{center}
\includegraphics[width=0.65\textwidth]{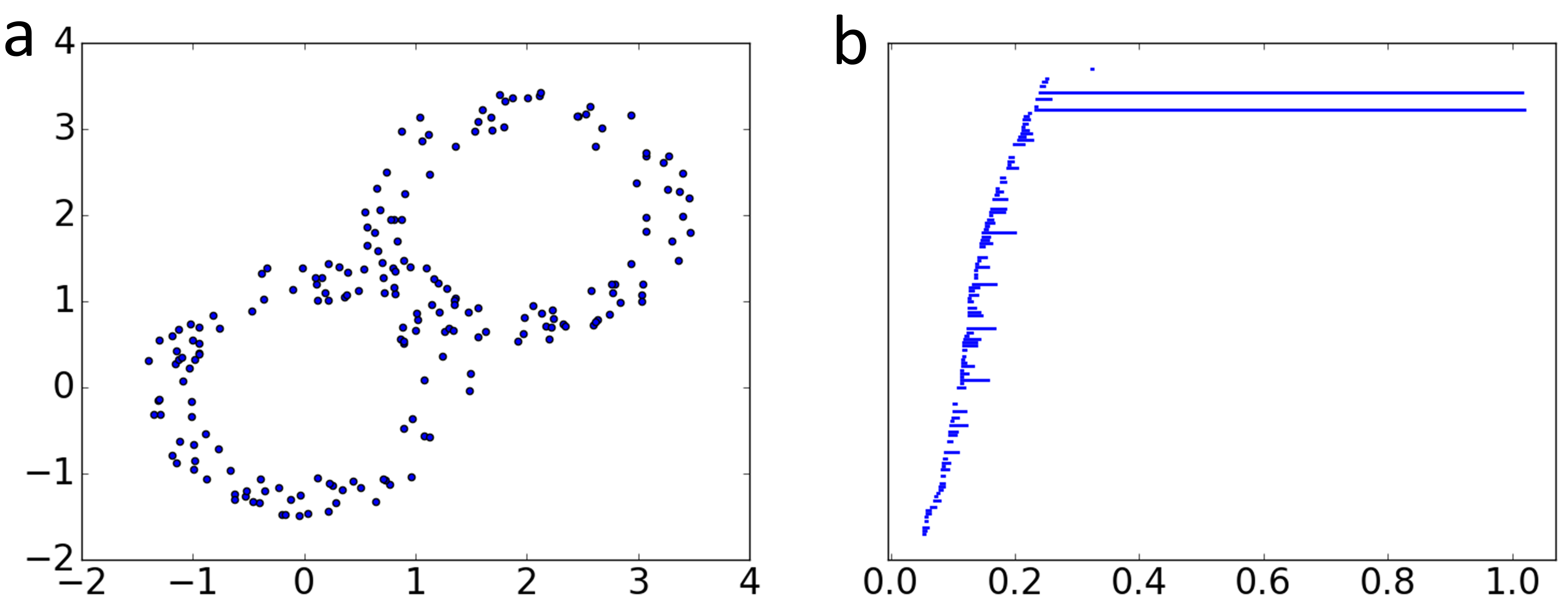}
\caption{\textbf{a}: A point cloud sampled from two adjacent annulus. \textbf{b}: The corresponding $H_1$ barcode using alpha complex filtration.}\label{fig:twoannulus}
\end{center}
\end{figure}

\begin{figure}[h]
\begin{center}
\includegraphics[width=0.65\textwidth]{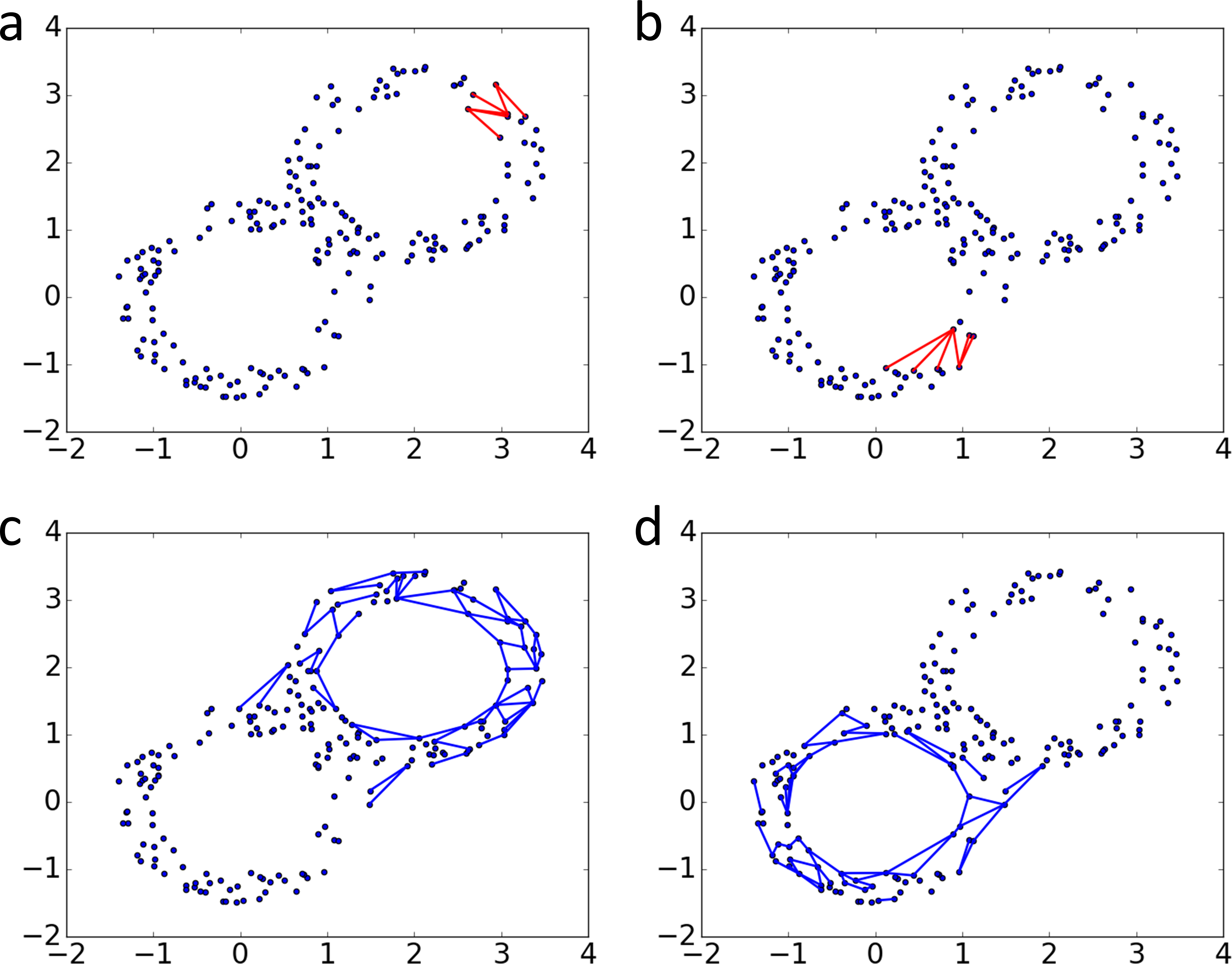}
\caption{\textbf{a} and \textbf{b}: Two representative $1$-cocycles corresponding to the two long $H_1$ bars. The edges where the cocycles take nonzero values are drawn in red. \textbf{c} and \textbf{d}: The smoothed $1$-cocycles associated to the representative cocycles. The edges where the cocycles take values with magnitudes greater than or equal to $0.035$ are drawn in blue. The smoothing is done on the subcomplexes associated to the filtration values at the middle of the corresponding bars.}
\label{fig:twoannulus_cocycles}
\end{center}
\end{figure}

We first consider a point cloud sampled from two adjacent annulus with radii $1$ and centered at $(0, 0)$ and $(2, 2)$ as shown in Figure~\ref{fig:twoannulus}. There are two significant $H_1$ bars associated to the two major circles. An example of the representative cocycles for the two long $H_1$ bars is shown in Figure~\ref{fig:twoannulus_cocycles}a and b. The associated smoothed cocycles obtained by using the method described in Section~\ref{sec:epb} are shown in Figure~\ref{fig:twoannulus_cocycles}c and d.

\begin{figure}[h]
\begin{center}
\includegraphics[width=0.65\textwidth]{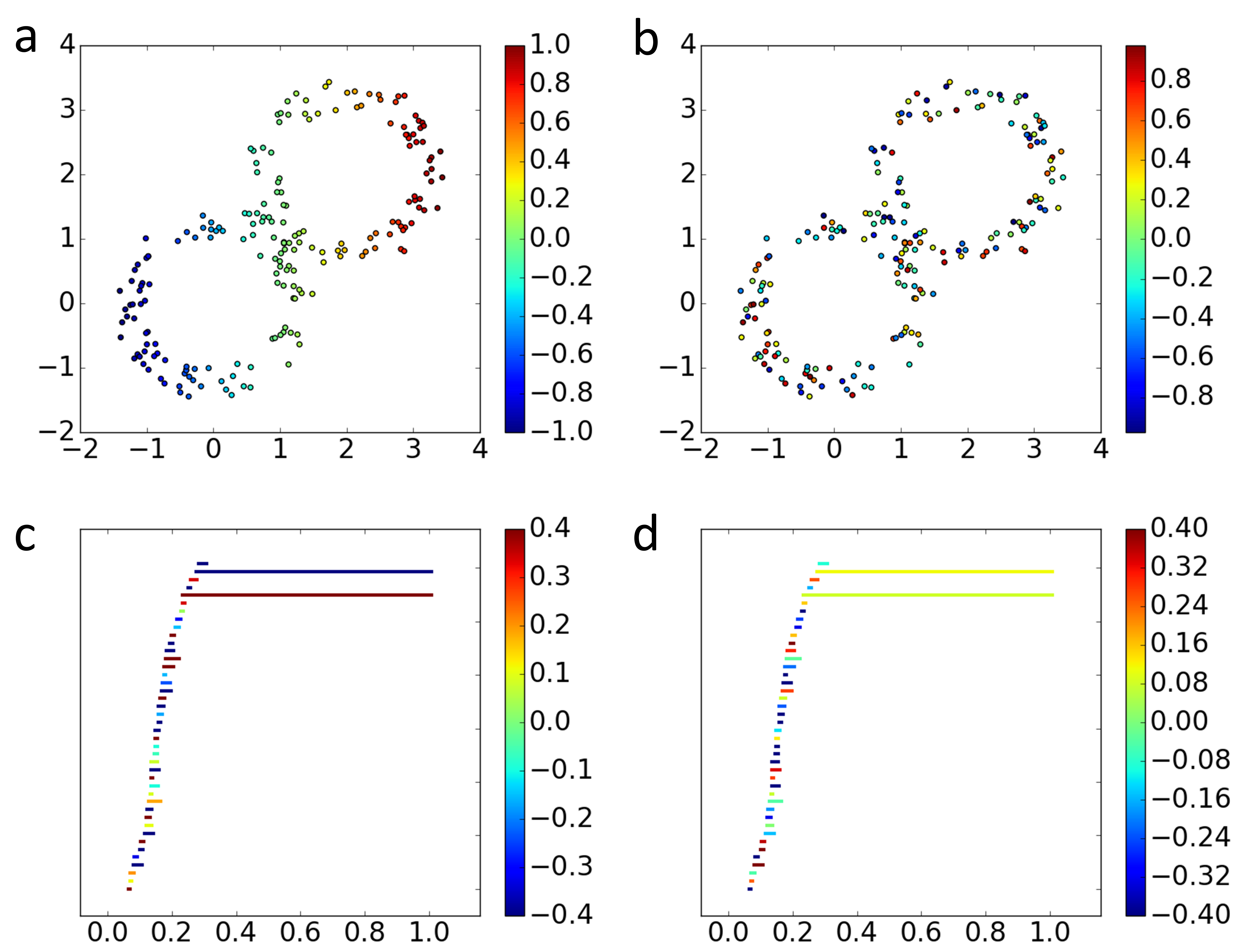}
\caption{\textbf{a} and \textbf{b}: Two datasets with similar geometry but different information given on the nodes. \textbf{c} and \textbf{d}: The differences are revealed in the   persistent cohomology enriched $H_1$ barcodes.}
\label{fig:colorbar}
\end{center}
\end{figure}

Given two datasets with similar geometry but different values on the nodes, we can use enriched barcodes to distinguish between them. See Figure~\ref{fig:colorbar} for example.

\FloatBarrier

\subsubsection*{Cuboid minus two balls}
\begin{figure}[h]
\begin{center}
\includegraphics[width=0.85\textwidth]{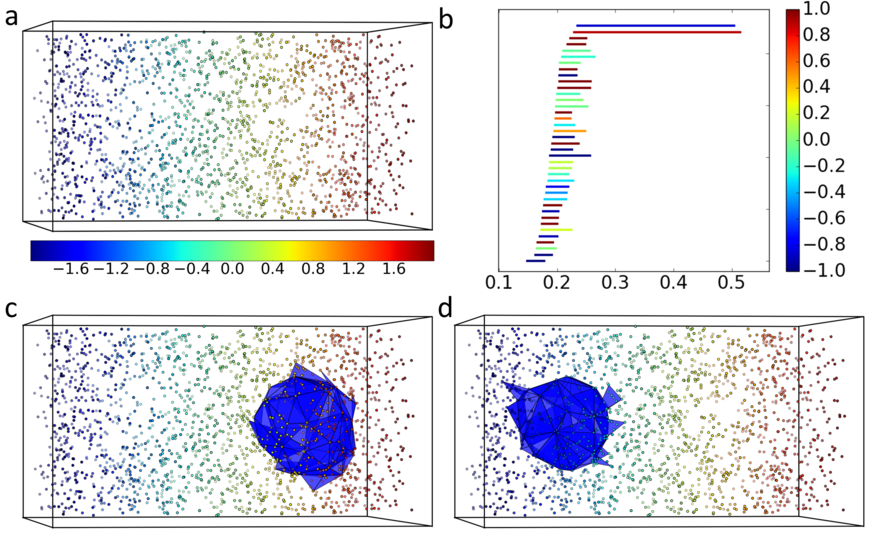}
\caption{\textbf{a}: The points sampled from an object which is a box subtracted by two balls. \textbf{b}: The   persistent cohomology enriched  $H_2$ barcode showing the two voids in the blue and red regions of the original dataset. \textbf{c} and \textbf{d}: The two smoothed $2$-cocycles. The faces where the cocycles take absolute values greater than or equal to $0.01$ are drawn in blue. The smoothing is done on the subcomplexes associated to the filtration values at the middle of the corresponding bars.}\label{fig:boxminustwoballs}
\end{center}
\end{figure}

In this example, the object considered is a rectangular cuboid ($[0,4]\times[0,2]\times[0,2]$) subtracted by two balls with radius of $0.5$ centered at $(1,1,1)$ and $(3,1,1)$. Two thousand points are first sampled from a uniform distribution over the cuboid and the ones that are inside the balls are deleted. The dataset with values on the points, the two smoothed cocycles corresponding to the two voids, and the enriched barcodes are shown in Figure~\ref{fig:boxminustwoballs}.

\subsection{Wasserstein distance based similarity}
\FloatBarrier

\begin{figure}
\begin{center}
\includegraphics[width=0.8\textwidth]{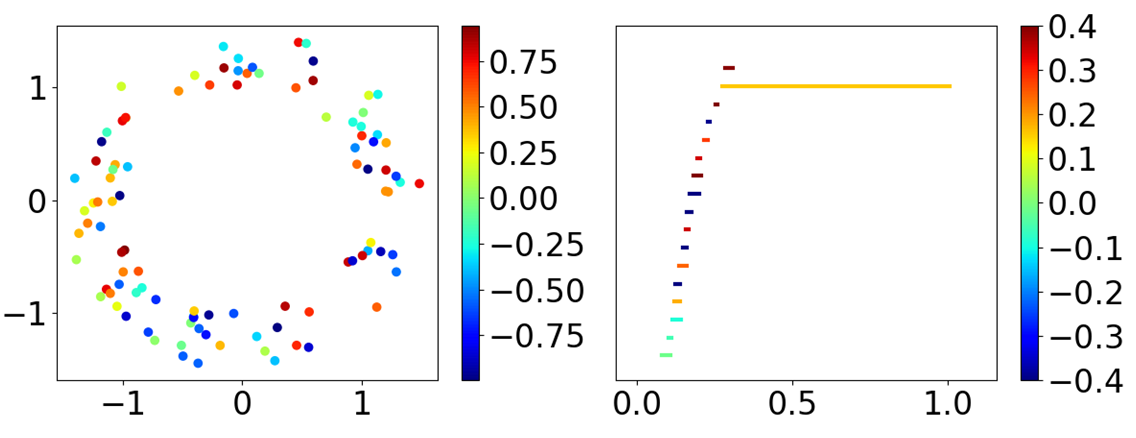}
\end{center}
\caption{D3 data sampled from an annulus with randomly assigned values on the points and corresponding  $H_1$ enriched barcode.}\label{fig:onecircle}
\end{figure}

\begin{figure}
\begin{center}
\includegraphics[width=0.5\textwidth]{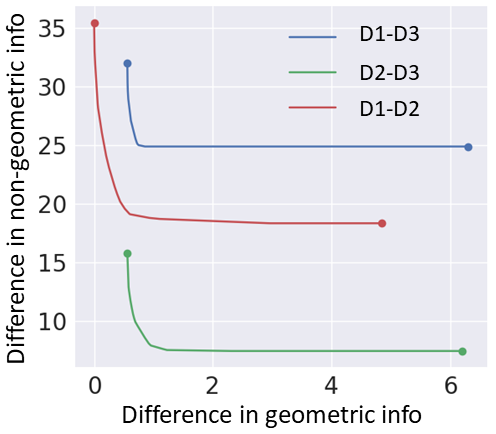}
\end{center}
\caption{Wasserstein characteristics curve.}\label{fig:wassersteincurve}
\end{figure}

We illustrate in this section the measurement of similarities among the persistent cohomology enriched barcodes. We use the enriched barcodes from three datasets, D1, D2, and D3. Here, D1 and D2 are the two datasets shown in Fig.~\ref{fig:colorbar}a and b, respectively, while  D3 is shown in the left chart of Fig.~\ref{fig:onecircle}. 
The Wasserstein characteristics curve defined in Eq. (\ref{eq:wassersteincurve}) for three datasets, i.e., D1, D2 and D3, are shown in Fig.~\ref{fig:wassersteincurve}. Here, D1 and D2 have the same geometry and thus their curve is more on the left side which means a smaller distance between their persistent homology barcodes. On the other hand, D3 has a similar value assignment on the points as that of D2, so their curve is on the bottom which means a smaller distance in the non-geometric information.

\FloatBarrier

\subsection{Persistent cohomology analysis of molecules}

Cyclic and cage-like structures often exist in complicated macromolecules in various scales. They can be as small as a benzene (a ring) containing  $6$ heavy atoms or an adamantane (a cage) containing $10$ heavy atoms. Some macromolecules have a global configuration of cyclic or cage-like structures such as buckminsterfullerene and carbon nanotubes which are consisted of tens or hundreds of atoms. Persistent cohomology is good at detecting these structures in multiple scales and when we label the atoms by their element types, we can also reveal the element composition of the detected structures. Specifically, if oxygen is of interest, we construct an input function $f_0$ (see Section~\ref{sec:inputfunction}) that is defined on the nodes representing the atoms, and outputs $1$ on oxygen atoms and $0$ elsewhere. We illustrate this application using a cyclic structure cucurbit[8]uril and a cage-like structure B$_{24}$N$_{24}$ cage in this section.

\subsubsection*{Cucurbituril}
\begin{figure}
\begin{center}
\includegraphics[width=0.8\textwidth]{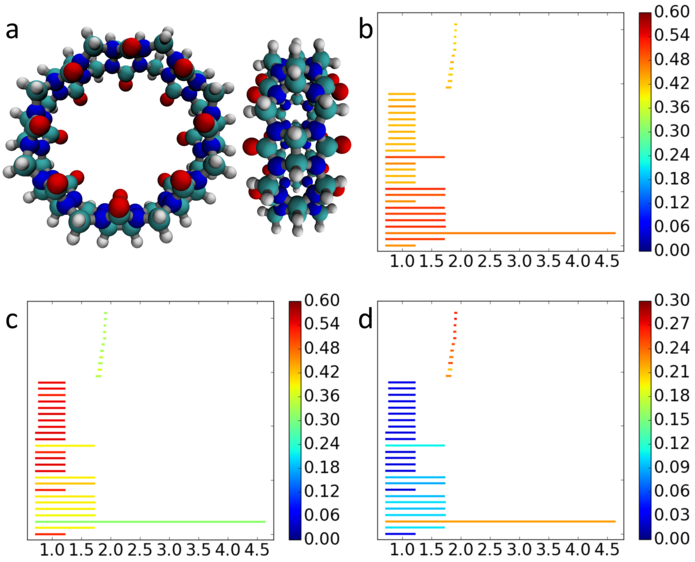}
\caption{\textbf{a}: The cucurbit[8]uril molecule viewed from two different angles. The hydrogen, carbon, nitrogen, and oxygen atoms are colored in white, grey, blue, and red, respectively. \textbf{b}, \textbf{c}, and \textbf{d}: The The   persistent cohomology enriched $H_1$ barcodes obtained by assigning $1$ to nodes of the selected atom types (carbon, nitrogen, and oxygen respectively) and $0$ elsewhere.}\label{fig:cb8}
\end{center}
\end{figure}

In this example, we consider a macrocyclic molecule cucurbit[8]uril from the cucurbituril family. The molecule contains eight $6$-membered rings and sixteen $5$-membered rings consisted of carbon and nitrogen atoms. The rings form a big cyclic structure with a relatively tighter opening surrounded by oxygen atoms. The structure is taken from the provided structure in the SAMPL6 challenge \cite{SAMPL6} and the resulting $H_1$ barcodes are shown in Figure~\ref{fig:cb8}.

\subsubsection*{Boron nitride cage}
\begin{figure}
\begin{center}
\includegraphics[width=0.8\textwidth]{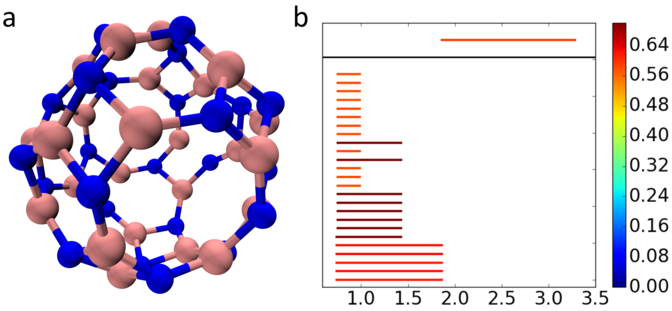}
\caption{\textbf{a}: A structure of the B$_{24}$N$_{24}$ cage. The nitrogen and boron atoms are colored in blue and grey. \textbf{b}: The enriched barcodes obtained by assigning $1$ to Boron atoms and $0$ elsewhere. The   persistent cohomology enriched $H_1$ and $H_2$ barcodes are plotted in bottom and top panels.}\label{fig:bncage}
\end{center}
\end{figure}

The fullerene-like boron nitride cages exhibit spherical structures similar to fullerenes but are consisted of boron and nitrogen atoms. The global spherical structure is composed of a collection of local rings containing several atoms. A possible structure of B$_{24}$N$_{24}$ cage given in the supporting information of \cite{zhang2013three} is used in this example. The molecule and the enriched barcode is shown in Figure~\ref{fig:bncage}.

\FloatBarrier

In this application, the element type could be substituted by other information that the user is interested in, such as partial charge, van der Waals potential, and electrostatic solvation free energy.

\subsection{Prediction of protein-ligand binding affinities}

An important component of computer-aided drug design is the prediction of protein-ligand binding affinities based on given protein-ligand complex structures. Persistent homology is good at identifying rings, tunnels, and cavities in various scales which are crucial to the protein-ligand complex stability and instability. In addition to geometry and topology, chemical and biological complexity is also needed to be addressed toward a practically useful method for this application. The important chemical and biological information includes atom properties such as atom types, atomic charges, interaction strengths and information from bioinformatics study such conservation scores of protein residues. To this end, for example, the behavior of atoms of different element types can be described by computing persistent homology for subsets of atoms of the molecule of certain element types \cite{ZXCang:2017b}. The interaction between protein and ligand can be emphasized by prohibiting an edge to form between two atoms both in the protein or the ligand. The electrostatic interactions can be revealed by tweaking the distance matrix used for filtration to be the interaction strength computed with a chosen physical model such as Coulomb's law \cite{ZXCang:2018a}. However, the approaches described above disturb the original geometry and topology of the protein-ligand complexes. With the method proposed in this work, we are able to naturally embed the information such as atom type, atomic partial charges, and electrostatic interactions to the barcodes without disturbing the original geometric and topological setup of the molecular systems.

\begin{figure}
\begin{center}
\includegraphics[width=0.8\textwidth]{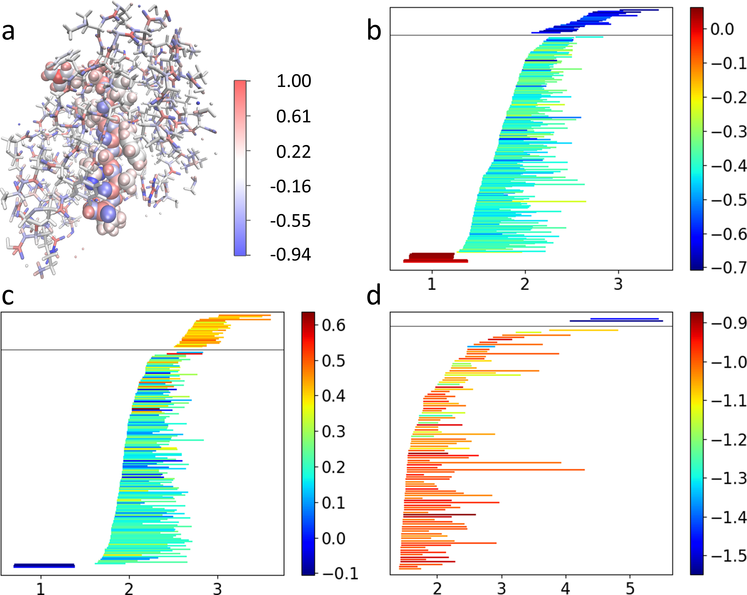}
\caption{Enriched barcodes focusing on atomic partial charges. \textbf{a}: Ligand (as van der Waals spheres) and surrounding protein atoms (within 12 \AA of ligand as thick sticks) of PDB entry 1a94. The color reflects the atomic partial charges. \textbf{b}, \textbf{c}, and \textbf{d}: Enriched barcodes for partial charges generated by computing persistent cohomology with alpha complex filtration on all heavy atoms, all carbon atoms, and nitrogen and oxygen atoms respectively. In each barcode subfigure, the top panel shows the   persistent cohomology enriched $H_2$ barcode and the bottom one shows the   persistent cohomology enriched  $H_1$ barcode.}\label{fig:proteinexample}
\end{center}
\end{figure}

We compute the persistent cohomology enriched barcodes for protein-ligand complexes, turn them into structured features, and combine with machine learning methods for the prediction of binding affinities. The procedure is validated on datasets from the PDBbind database \cite{PDBBind:2015} which includes experimentally derived protein-ligand complex structures and the associated binding affinities. An example of enriched barcode for atomic partial charges is shown in Fig.~\ref{fig:proteinexample}.

\subsubsection*{Enriched barcodes generation}

In addition to the traditional barcode obtained from persistent homology computation, we would also like to add descriptions of the electrostatic properties of the system. An efficient characterization of this property is the Coulomb potential where the interaction between two point charges is relatively described by $q_iq_j/r_{ij}$ where $q_i$ and $q_j$ are the point charges with a distance of $r_{ij}$. The atomic partial charges of proteins are assigned by using PDB2PQR software \cite{Pdb2pqr} with CHARMM22 force field. Two types of construction of the physical information are used to characterize the systems.

For dimension $0$, a collection of subsets of atoms is first identified according to atom types. Specifically, 10 element types (C, N, O, S, P, F, Cl, Br, I, H) are considered for ligands and 5 element types are considered for proteins (C, N, O, S, H) and a total of 50 subsets of atoms are selected by choosing one element type from each component (protein or ligand). The pairwise distance matrix based on Euclidean distance is tweaked by setting distances between atoms both from protein or ligand to infinity which emphasizes the interactions between protein and ligand. Based on the tweaked distance matrix, persistent (co)homology computation with Rips complex is performed. The electric potential is computed for each atom with its nearest neighbor in the different part of the protein-ligand complex and is put on this atom as the additional information. We define the input function $f_0^0: X^0\rightarrow \mathbb{R}$ to take $0$ on protein atoms and to take the value discussed above on ligand atoms. The average potential over ligand atoms in each $0$-cocycle representative is used to generate features. In this way, the favorability of the protein ligand electrostatic interactions is explicitly described.

For dimensions $1$ and $2$, the input function $f_0^1: X^1\rightarrow\mathbb{R}$ is defined to output the absolute value of electric potential on edges connecting two atoms to characterize the interaction strengths. The Coulomb potential is modeled as
\begin{equation*}
E_{ij} = k_e\frac{q_iq_j}{r_{ij}},
\end{equation*}
where $k_e$ is Coulomb's constant, $q_i$ and $q_j$ are the partial charges of atoms $i$ and $j$, and $r_{ij}$ is the distance between the two atoms.
Persistent (co)homology with alpha complex is computed on three subsets of the protein-ligand complexes, all heavy atoms, all carbon atoms, and all oxygen/nitrogen atoms. For simplicity, all enriched barcodes are computed only at the middle points of the bars.

\subsubsection*{Featurization of barcodes}
Given an enriched barcode, $B=\{\{b_i,d_i,f^*_i\}\}_{i\in I}$ obtained by applying the proposed method to a dataset with an input function $f_0$ (see Section~\ref{sec:inputfunction}), we turn it into fixed shape array required by the machine learning algorithms we choose. Here, the input function is $f_0^0$ or $f_0^1$ described in the previous section when computing $0$th dimensional persistent (co)homology or in higher dimensions.

For dimension $0$, we first identify a range of scales to focus on and in this application, we are interested in the interval $[0,12)\mbox{\AA}$. The interval is then divided into $6$ subintervals $\{[l^0_j, r^0_j)\}_j=\{[0,2.5),\allowbreak [2.5,3),\allowbreak [3,3.5),\allowbreak [3.5,4.5),\allowbreak [4.5,6),\allowbreak [6,12)\}$ to address different types of interactions. For dimension $0$, we are interested in the death values of the bars. Therefore, a collection of index sets marking the deaths values of the bars that fall into each subinterval is calculated as
\begin{equation*}
I^0_j = \{i\in I|l_j\leq d_i <r_j\}.
\end{equation*}

For dimensions $1$ and $2$,  we are interested in the interval $[0,6)\mbox{\AA}$ with Alpha complex filtration. The interval is then divided into $6$ equal length subintervals $\{[l^{1,2}_j,r^{1,2}_j)\}_j$. 
We then define a collection of index sets marking the bars that overlap with each subinterval,
\begin{equation*}
I^{1,2}_j = \{i\in I|b_i <r^{1,2}_j, d_i \geq l^{1,2}_j\}.
\end{equation*}

Given a collection of index sets $\{I_j\}_j$, a feature vector $\mathbf{v}^h(B)$ is defined as
\begin{equation*}
\left(\mathbf{v}^h(B)\right)_j = |I_j|.
\end{equation*}
When $\{I^0_j\}_j$ is used, it characterizes the number of component merging events in each filtration parameter interval. When $\{I^{1,2}_j\}$ is used, it reflects the ranks of homology groups at certain stage along the course of filtration.

A feature vector $\mathbf{v}^f(B,f_0)$ can be generated subsequently to address the information of the predefined function on the homology generators,
\begin{equation*}
\left(\mathbf{v}^f(B,f_0)\right)_j = \frac{\sum\limits_{i\in I_j}\bar{f}^*_i}{|I_j|},
\end{equation*}
where $\bar{f}^*_i = (\int_{b_i}^{d_i}f^*_i(x)dx)/(d_i-b_i)$.

\subsubsection*{Machine learning algorithm}

The application of predicting protein-ligand binding affinity based on structures can be regarded as a supervised learning problem. Generally speaking, we are given a collection of pairs of input and output $\{(x_i,y_i)\}$ and there is a chosen model which is a function $M(x;\theta)$ with tunable parameters $\theta$. The training process is to find a specific setting for the function $M$ that globally or locally minimizes a penalty function which depends on the given data $\{(x_i,y_i)\}$ and the parameter set $\theta$. Once trained, the model can be used to predict the output for a newly given input.

In general, the proposed persistent cohomology can be combined with any advanced machine learning algorithm, such as deep neural networks used in our earlier work \cite{ZXCang:2017c}. However, the goal of the present work is to illustrate the utility of the proposed  persistent cohomology. Therefore, we choose a basic algorithm, the  gradient booting trees (GBT) method,  for testing the accuracy, robustness, and efficiency of our method. GBT is an ensemble of trees method with single decision trees as building blocks. The training of a GBT model is done by adding one tree at a time according to reduce loss of current model. In practice, different randomly selected subsets of the training data and features are used for each update of the model to reduce overfitting. For every result reported in Table~\ref{tab:pdbbind}, a parameter search is done by cross-validation within the training set where model performance is judged by Pearson's correlation coefficient. The candidate values for hyper-parameters tried are summarized in Table~\ref{tab:candidateprm}. Another hyper-parameter max\_feature is set to sqrt because of the relatively large number of features. The GradientBoostingRegressor module in scikit-learn (version 0.17.1) \cite{scikit-learn} software is used.

\begin{table}
\begin{center}
\begin{tabular}{|l|l|}
\toprule
Hyper Prm. & Candidates \\
\midrule
\midrule
n\_estimators & 5000, 10000, 20000 \\
\midrule
max\_depth & 4, 8, 16 \\
\midrule
min\_samples\_split & 5, 10, 20 \\
\midrule
learning\_rate & 0.0025, 0.005, 0.01 \\
\midrule
subsample & 0.25, 0.5, 0.75 \\
\midrule
min\_samples\_leaf & 1, 3 \\
\bottomrule
\end{tabular}
\end{center}
\caption{Candidate values for hyper-parameters of the gradient boosting trees model.}\label{tab:candidateprm}
\end{table}

\subsubsection*{Binding affinity predictions}

\begin{table}
\begin{center}
\begin{tabular}{|c|c|c|c|c|}
\toprule 
 PDBbind & v2007 & v2013 & v2015 & v2016 \\ 
\midrule 
\midrule 
Dim $0$  PH & 0.802 (1.47)) & 0.754 (1.56) & 0.745 (1.56) & 0.824 (1.32) \\ 
\midrule
Dim $0$ PC & 0.796 (1.50) & 0.768 (1.53) & 0.763 (1.53) & 0.833 (1.31) \\ 
\midrule
\midrule
Dim $1$\&$2$ PH & 0.726 (1.65) & 0.706 (1.67) & 0.718 (1.62) & 0.767 (1.46) \\ 
\midrule
Dim $1$\&$2$ PC & 0.738 (1.64) & 0.734 (1.60) & 0.737 (1.59) & 0.778 (1.44) \\ 
\bottomrule

\end{tabular}
\caption{The predictor performance is evaluated by training on PDBbind refined set excluding the core set and testing on the core set of a certain year's version. The median Pearson's correlation coefficient (root mean squared error in pKd/pKi unit) among 10 repeated experiments is reported for persistent homology (PH) and persistent cohomology (PC). In the PC, electrostatic information is utilized. }\label{tab:pdbbind}
\end{center}
\end{table}
We test the improvement of the enriched barcodes with electrostatic information in the cases of $0$th dimension and higher dimensions using the PDBbind database. The predictor performance is improved by using the enriched barcode embedding the electrostatics information. The results are listed in Table~\ref{tab:pdbbind}, which are near optimal. This study indicates that the electrostatic information incorporated in the persistent cohomology improves the binding affinity predictions except for the case of PDBbind v2007. The approach proposed in this work can be generalized to other physical properties, such van der Waals interactions.   

\section{Conclusion}

Algebraic topology, particularly persistent homology has been devised to simplify high-dimensional complex geometric information in terms of topological invariants. However, during the topological abstraction of biomolecular datasets   some physical, chemical and biological information  is neglected. Therefore,  there is a  pressing need to embed physical, chemical and biological information, such as atom types, partial charges, and pairwise interaction strengths in a dataset into the topological invariants generated from the geometric (i.e., structural) information of the dataset.  In general, when analyzing datasets with persistent homology, the geometric information is built into topological invariants while non-geometric information is usually neglected. In duality to homology, cohomology allows us to retain crucial non-geometric information  in topological modeling. Utilizing the richer information carried by cohomology, we introduce a framework to encode  in the topological invariants the additional physical information from the dimensions that are not used for persistent homology computation. The non-geometric information is attached to the topological invariants in regular persistent homology computation. This is achieved by finding a smoothed representative cocycle with respect to a Laplacian directly defined on the simplicial complexes or a weighted graph Laplacian. The smoothed cocycles then serve as measures on the simplicial complexes and allow us to integrate  the additional information. As a result, in addition to the original persistence barcodes, functions of filtration values associated to each persistence pair are constructed, which enriches the information carried by the original barcodes. A similarity score based on Wasserstein distance is introduced to analyze these enriched barcodes. The proposed method is applied to the protein-ligand binding affinity prediction that motivated the current development. We show that   by adding electrostatics information to the barcodes,  the present persistent cohomology improves the performance in the practical prediction of protein-ligand binding affinities. The results obtained from the proposed method are near optimal for these datasets.

The proposed method is potentially useful for a wide range of applications where data come with multiple heterogeneous dimensions including the geometric dimensions as a subset. 
For one specific dimension of a multidimensional dataset, there are also cases where we would like to embed the information carried in this dimension to the persistence barcodes computed for other dimensions rather than looking at the persistence for this dimension. For example, persistent homology can find us loops and voids in biomolecular structures and we are interested in question that what kind of physical properties (such as charges and interactions ) do these homology generators carry. In this case, the duality between homology and cohomology enables us to better localize the homology generators and to examine the physical/biological  interactions and functions  associated to each generator.

 \section*{Acknowledgments}
This work was supported in part by NSF grants  IIS-1302285, DMS-1721024 and DMS-1761320. 
\FloatBarrier

\end{document}